\documentclass[prb,aps,twocolumn,superscriptaddress,longbibliography]{revtex4-2}
\usepackage{graphicx,color}
\usepackage{amsthm}
\usepackage{amsfonts}
\usepackage{algorithmic}
\usepackage{enumerate}
\usepackage{latexsym}
\usepackage{amsmath}
\usepackage{amssymb}
\usepackage{multirow}
\usepackage{array}
\usepackage{diagbox}
\usepackage{subfigure}
\usepackage[colorlinks=true,citecolor=blue,linkcolor=blue]{hyperref}
\usepackage{mhchem} 
\usepackage[normalem]{ulem} 

\emergencystretch=\maxdimen
\makeatletter
\DeclareFontFamily{OMX}{MnSymbolE}{}
\DeclareSymbolFont{MnLargeSymbols}{OMX}{MnSymbolE}{m}{n}
\SetSymbolFont{MnLargeSymbols}{bold}{OMX}{MnSymbolE}{b}{n}
\DeclareFontShape{OMX}{MnSymbolE}{m}{n}{
    <-6>  MnSymbolE5
   <6-7>  MnSymbolE6
   <7-8>  MnSymbolE7
   <8-9>  MnSymbolE8
   <9-10> MnSymbolE9
  <10-12> MnSymbolE10
  <12->   MnSymbolE12
}{}
\DeclareFontShape{OMX}{MnSymbolE}{b}{n}{
    <-6>  MnSymbolE-Bold5
   <6-7>  MnSymbolE-Bold6
   <7-8>  MnSymbolE-Bold7
   <8-9>  MnSymbolE-Bold8
   <9-10> MnSymbolE-Bold9
  <10-12> MnSymbolE-Bold10
  <12->   MnSymbolE-Bold12
}{}

\let\llangle\@undefined
\let\rrangle\@undefined
\DeclareMathDelimiter{\llangle}{\mathopen}%
                     {MnLargeSymbols}{'164}{MnLargeSymbols}{'164}
\DeclareMathDelimiter{\rrangle}{\mathclose}%
                     {MnLargeSymbols}{'171}{MnLargeSymbols}{'171}
\makeatother

\NewDocumentCommand{\dgal}{sO{}m}{%
  \IfBooleanTF{#1}
    {\dgalext{#3}}
    {\dgalx[#2]{#3}}%
}
\NewDocumentCommand{\dgalext}{m}{%
  \sbox0{%
    \mathsurround=0pt 
    $\left\{\vphantom{#1}\right.\kern-\nulldelimiterspace$%
  }%
  \sbox2{\{}%
  \ifdim\ht0=\ht2
    \{\kern-.625\wd2 \{#1\}\kern-.625\wd2 \}%
  \else
    \left\{\kern-.7\wd0\left\{#1\right\}\kern-.7\wd0\right\}%
  \fi
}

\begin{document}
\title{Interplay of magnetic and thermodynamic responses in the kagome-triangular system}
\author{Zixuan Jia}
\affiliation{School of Physics and Astronomy, and Key Laboratory of Multiscale Spin Physics (Ministry of Education), Beijing Normal University, Beijing 100875, China}
\author{Lufeng Zhang}
\affiliation{School of Physical Science and Technology, Beijing University of Posts and Telecommunications, Beijing 100876, China}
\author{Qingzhuo Duan}
\affiliation{School of Physics and Astronomy, and Key Laboratory of Multiscale Spin Physics (Ministry of Education), Beijing Normal University, Beijing 100875, China}
\author{Zenghui Fan}
\affiliation{School of Physics and Astronomy, and Key Laboratory of Multiscale Spin Physics (Ministry of Education), Beijing Normal University, Beijing 100875, China}
\author{Jingyao Wang}
\affiliation{Key Laboratory for Microstructural Material Physics of Hebei Province, School of Science, Yanshan University, Qinhuangdao 066004, China}
\author{Ying Liang}
\affiliation{School of Physics and Astronomy, and Key Laboratory of Multiscale Spin Physics (Ministry of Education), Beijing Normal University, Beijing 100875, China}
\author{Bing Huang}
\affiliation{School of Physics and Astronomy, and Key Laboratory of Multiscale Spin Physics (Ministry of Education), Beijing Normal University, Beijing 100875, China}
\affiliation{Beijing Computational Science Research Center, Beijing 100084, China}
\author{Tianxing Ma}
\email{txma@bnu.edu.cn}
\affiliation{School of Physics and Astronomy, and Key Laboratory of Multiscale Spin Physics (Ministry of Education), Beijing Normal University, Beijing 100875, China}

\begin{abstract}  
Inspired by the recent experimental progress in pyrochlore derivative \ce{RE3Sb3A2O14 (A=Mg, Zn)}, 
we investigate the Hubbard model on the kagome lattice with an additional hopping $t'/t$, which enables continuous interpolation between the kagome and triangular lattices by using determinant quantum Monte Carlo simulations. 
We find that increasing $t'/t$ suppresses the nearest-neighbor antiferromagnetic correlations. Concurrently, the next-nearest-neighbor antiferromagnetic correlations are enhanced and closely associated with the emergence of a pronounced low-temperature peak in the specific heat.
Increasing on-site interaction $U$ enhances magnetic correlations and shifts the associated $t'/t$ crossover points to larger values. 
We also discuss the sign problem to clarify which parameter region of our numerical simulations is accessible and reliable. 
Our results uncover the competition between frustration and correlations and the interplay of magnetic and thermodynamic responses in the kagome lattice, providing insights into correlated states in frustrated materials.
\end{abstract}
\maketitle

\section{Introduction}
The kagome lattice has attracted considerable attention in strongly correlated electron systems. 
The electronic band structure of the kagome lattice is characterized by three key features: a flat band at the high-energy edge, a Dirac cone of the lower bands, and multiple van Hove singularities arising from the saddle points in the band dispersion \cite{Ohgushi2000,Bergman2008}.
The interplay of these features gives rise to nontrivial electronic topology, correlated many-body states, and exotic magnetic phases \cite{Syôzi1951,Kang2020,Lin2018,Yin2019,Ortiz2019,Jiang2022,YangCPB2024}.
The kagome lattice is also considered a promising platform for realizing quantum spin liquid states, although the underlying mechanism remains debated \cite{Han2022,Mingxuan2015,Norman2016}.
Over the past decades, remarkable magnetic phases have been observed in various kagome materials, including herbertsmithite compounds \cite{Helton2007,Mendels2007,Helton2010} and pyrochlore derivatives \cite{Dun2016,Scheie2016,Dun2017,Vinod2022}. It thus serves as an interesting platform for studying magnetic phases in frustrated systems.

To understand the origin of its exotic magnetic phases, the kagome lattice has been extensively studied theoretically. 
For the nearest-neighbor spin-1/2 Heisenberg antiferromagnet, geometric frustration and quantum fluctuations favor a disordered ground state. 
Numerical studies employing methods such as density matrix renormalization group, tensor network approaches, and variational Monte Carlo have yielded results that differ to some extent, leaving the nature of the ground state unclear \cite{Yan2011,Depenbrock2012,He2017}.
Extended Heisenberg models, including $J_1$-$J_2$-$J_3$ interactions, further reveal a rich landscape of competing phases, potentially leading to valence-bond solid order or weak magnetic order \cite{Messio2012,Gong2015,Tsezar2015}.
The Hubbard model provides a complementary perspective by capturing the competition between electronic itinerancy and electron interaction. 
As the interaction strength increases, the Hubbard model on the kagome lattice undergoes a Mott insulating transition accompanied by unconventional magnetic responses \cite{Medeiros-Silva2023,Duan2025}. 
The flat band significantly amplifies correlation effects, giving rise to ferromagnetism and topological phases such as Chern insulators \cite{Mielke1991,Tasaki1992,Medeiros-Silva2023,Kiesel2012,Lima2023}.
Comparative studies between the kagome and triangular lattices provide valuable insights due to their similar geometric structures and strong frustration \cite{Berlinsky1994}.
In the spin-1/2 Heisenberg model, the triangular lattice exhibits a magnetically ordered ground state, whereas the kagome lattice is generally considered disordered \cite{Diep2012}. 
Quantum Monte Carlo studies of the Hubbard model further show that the doped kagome lattice hosts stronger short-range magnetic correlations than the triangular lattice \cite{Bulut2005}.
Although the kagome and triangular lattices share certain magnetic features, the kagome lattice exhibits special magnetic behavior due to its distinctive density of states and enhanced correlation effects.
Despite extensive theoretical efforts, the ground-state properties, global phase diagram, and mechanisms of competing interactions in the kagome lattice remain incompletely understood, underscoring the need for further investigation.

Meanwhile, the thermodynamic properties of the kagome lattice remain a subject of considerable interest. 
Within the Heisenberg model, both the kagome lattice and the triangular lattice exhibit a single-peak structure in the specific heat \cite{Wang1992}. 
In contrast, dynamical mean-field theory studies of the Hubbard model suggest that the kagome lattice develops a small secondary peak in the specific heat, attributed to spin-chirality degrees of freedom \cite{Udagawa2010}. 
Unbiased quantum Monte Carlo simulations of the Hubbard model indicate that the specific heat of the kagome lattice consistently retains a broad single peak across interaction strengths, with a subtle low-temperature shoulder feature \cite{Medeiros-Silva2023}. 
Moreover, continuous interpolation between the kagome and triangular lattices has been studied using a modified Heisenberg model with two exchange couplings $J$ and $\eta J$. 
It has revealed that both limiting cases display single-peak behavior, while for $\eta \ll 1$ the model predicts a double-peak structure consistent with experimental observations of helium adsorbed on graphite \cite{Wang1992}. 
These results indicate that although some qualitative features of the thermodynamic properties of the kagome lattice are robust, its low-temperature behavior remains unclear. 
Additionally, systems with intermediate properties between the kagome and triangular lattices may exhibit unusual physical behaviors, which are worthy of further investigation.

Pyrochlore derivatives \ce{RE3Sb3A2O14 (A=Mg, Zn)} provide a promising platform for investigating the magnetic frustration inherent to the kagome lattice. 
These systems exhibit a wide range of magnetic behaviors, including long-range magnetic order, scalar chiral magnetic order, short-range spin correlations, spin-ice behavior, and the emergent charge order state \cite{Jiang2024,Dun2016,Scheie2016,Dun2017,Vinod2022}. 
However, RE/A site-mixing disorder is commonly observed in the \ce{RE3Sb3A2O14 (A=Mg, Zn)} family. 
It not only disrupts the ideal lattice geometry but also gives rise to intriguing physical phenomena. 
For instance, in \ce{Dy3Sb3Zn2O14} with $1\% \text{--} 5\%$ Dy/Zn site mixing, pronounced anomalies in both the specific heat and ac susceptibility have been observed at low temperatures, interpreted as signatures of a transition into the emergent charge order  state \cite{Dun2017}. 
Additional low-temperature anomalies in the specific heat have been reported, attributed to dipolar interactions or exchange interactions extending beyond the nearest neighbors, resembling the behavior of dipolar spin-ice systems \cite{Jiang2024}. 
The introduction of site-mixing disorder in \ce{Dy3Sb3Zn2O14} transforms the lattice into a diluted two-dimensional triangular geometry, or an intermediate structure between kagome and pyrochlore lattices. 
Therefore, synthesizing and studying such systems that bridge the kagome and triangular limits, including possible intermediate states, provide a valuable opportunity to explore their unconventional magnetic and thermodynamic properties.

\begin{figure}[tbp]
\includegraphics[width=0.8\columnwidth]{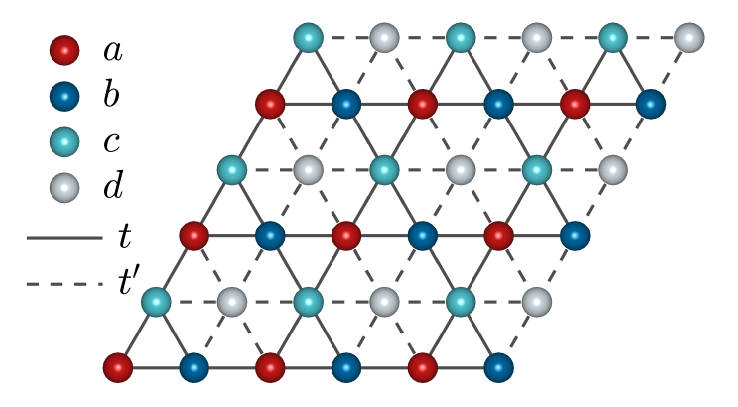}
\caption{
Sketch of the kagome lattice with additional $d$ sites. Points $a$, $b$, and $c$ form the kagome unit cell, and the $d$ sites are coupled to the kagome sublattice via the hopping $t'$. Solid and dashed bonds denote hopping amplitudes $t$ and $t'$. The system continuously interpolates between the kagome and triangular lattice limits by varying $t'/t$ from 0 to 1.}
\label{Fig1}
\end{figure}

In this study, we employ the determinant quantum Monte Carlo (DQMC) method to explore magnetic correlations and thermodynamic properties of the half-filled Hubbard model on the kagome lattice, introducing a tunable hopping parameter $t'/t$ as illustrated in Fig.~\ref{Fig1}. 
The system is continuously interpolated between the kagome and triangular lattices by varying $t'/t$ from 0 to 1.
We find that increasing $t'/t$ suppresses the nearest-neighbor spin-spin correlations. Meanwhile, the next-nearest-neighbor correlations are enhanced and closely associated with the emergence of a pronounced low-temperature peak in the specific heat.
Our results reveal a possible correlation between magnetic and thermodynamic responses in the frustrated kagome system \cite{Paiva2001,Paiva2005,Medeiros-Silva2023}.
The paper is structured as follows. Section~\ref{secmodel} presents the Hubbard model, key observables and the DQMC method. Section~\ref{secresults} analyzes the evolution of magnetic correlations and thermodynamic properties with varying $t'/t$ and interaction strength $U$. Section~\ref{secconclusions} summarizes the main conclusions and discusses implications for tuning frustration and correlations in the kagome system.

\section{Model and methods}\label{secmodel}
We consider the Hubbard model defined on a two-dimensional kagome lattice with a set of additional sites located on the $d$ sites as shown in Fig.~\ref{Fig1}. The Hamiltonian is defined by 
\begin{equation}
\label{eq1}
\begin{gathered}
\mathcal{H} = \mathcal{H}_1 + \mathcal{H}_2 + \mathcal{H}_3, \\[2ex]
\begin{aligned}
\mathcal{H}_1 & = -t\sum_{\mathbf{r},\sigma}
 \left( a_{\mathbf{r},\sigma}^\dagger b_{\mathbf{r},\sigma}
+ b_{\mathbf{r},\sigma}^\dagger c_{\mathbf{r},\sigma}
+ c_{\mathbf{r},\sigma}^\dagger a_{\mathbf{r},\sigma} \right. \\
& \quad \left. + a_{\mathbf{r},\sigma}^\dagger b_{\mathbf{r}-\mathbf{x},\sigma}
+ b_{\mathbf{r},\sigma}^\dagger c_{\mathbf{r}+\mathbf{x}-\mathbf{y},\sigma}
+ c_{\mathbf{r},\sigma}^\dagger a_{\mathbf{r}+\mathbf{y},\sigma} +\mathrm{H.c.} \right), \\[1ex]
\mathcal{H}_2 & = - t^{\prime}\sum_{\mathbf{r},\sigma}
 \left( b_{\mathbf{r},\sigma}^\dagger d_{\mathbf{r},\sigma}
+ c_{\mathbf{r},\sigma}^\dagger d_{\mathbf{r},\sigma}
+ a_{\mathbf{r},\sigma}^\dagger d_{\mathbf{r}-\mathbf{x},\sigma} \right. \\
& \quad \left. + c_{\mathbf{r},\sigma}^\dagger d_{\mathbf{r}-\mathbf{x},\sigma}
+ a_{\mathbf{r},\sigma}^\dagger d_{\mathbf{r}-\mathbf{y},\sigma}
+ b_{\mathbf{r},\sigma}^\dagger d_{\mathbf{r}-\mathbf{y},\sigma} + \mathrm{H.c.} \right), \\[1ex]
\mathcal{H}_3 & = U \sum_{\mathbf{r},\alpha} n_{\mathbf{r},\uparrow}^\alpha n_{\mathbf{r},\downarrow}^\alpha - \mu \sum_{\mathbf{r},\sigma,\alpha} n_{\mathbf{r},\sigma}^\alpha,
\end{aligned}
\end{gathered}
\end{equation}
where $t$ denotes the nearest-neighbor hopping on the kagome lattice and $t^{\prime}$ denotes the hopping between the kagome sites $a$, $b$, $c$ and the additional sites $d$. We set $t=1$ as the unit of energy. 
The fermionic operators $a_{\mathbf{r},\sigma}^\dagger (a_{\mathbf{r},\sigma}), b_{\mathbf{r},\sigma}^\dagger (b_{\mathbf{r},\sigma}), c_{\mathbf{r},\sigma}^\dagger (c_{\mathbf{r},\sigma})$ and  $d_{\mathbf{r},\sigma}^\dagger (d_{\mathbf{r},\sigma})$ are the creation (annihilation) of an electron with spin $\sigma=\uparrow,\downarrow$ at position $\mathbf{r}$, satisfying the anticommutation relations. The particle number operator $n_{\mathbf{r},\sigma}^\alpha$ gives the number of electrons, $\alpha$ = $a$, $b$, $c$, or $d$.
The Hubbard interaction $U$ represents the strength of the on-site repulsion and the chemical potential $\mu$ is used to tune the electron density of the system to half-filling.

We employ DQMC simulations \cite{White1989,santos2003} to investigate the finite-temperature properties of the Hubbard model defined in Eq.~(\ref{eq1}). 
The partition function $\mathcal{Z} = \mathrm{Tr} e^{-\beta \mathcal{H}}$ is expressed as a path integral discretized into imaginary-time slices $\Delta \tau$ by the Suzuki-Trotter decomposition, where $\beta = 1/(k_{\mathrm{B}}T)$. The imaginary-time step $\Delta \tau$ is chosen sufficiently small such that Trotter errors remain smaller than statistical sampling errors. 
The interaction term is then decoupled using the Hubbard-Stratonovich (HS) transformation, introducing standard Ising HS fields that depend on both spatial sites and imaginary time.
This procedure renders the Hamiltonian quadratic in fermion operators, allowing the fermionic trace to be evaluated analytically. As a result, the partition function reduces to the product of spin-up and spin-down fermion determinants. These determinants serve as Boltzmann weights for Monte Carlo sampling of the HS fields, enabling the evaluation of Green’s functions and related observables such as spin, charge, and pairing correlations. 
In practice, we employ the Metropolis algorithm to update the HS fields. 
For ordered systems at large $\beta U$, global updates of the HS fields are often considered in DQMC simulations \cite{Scalettar1991}. In our simulations, we include one global update attempt per Monte Carlo sweep.
Simulations are initialized with random HS field configurations, followed by 8000 warm-up sweeps to reach equilibrium. And 10000-30000 additional sweeps are made to generate measurement data. 
To reduce autocorrelation effects, the measurements are divided into 10 bins to obtain coarse-grained averages, and the errors are estimated from the standard deviations from these averages. 
To evaluate the reliability of our results against the infamous sign problem as the particle-hole symmetry is broken, a careful analysis of the average sign is conducted.

To study the magnetic properties of the system, we calculate the real-space spin-spin correlations \cite{Lima2023},
\begin{equation}
c^{\alpha \gamma}(\mathbf{r}) = \frac{1}{3} \langle \mathbf{S}_{\mathbf{r}_0}^{\alpha} \cdot \mathbf{S}_{\mathbf{r}_0 + \mathbf{r}}^{\gamma} \rangle,
\label{eq2}
\end{equation}
where $\mathbf{r}_0$ denotes the position of any lattice site, and $\mathbf{r}$ represents the distance between two sites. The indices $\alpha$ and $\gamma$ label the sites $a$, $b$, $c$ or $d$. We also evaluate the local moment, defined as
\begin{equation}
\langle m^2 \rangle = \langle \left( n_{\mathbf{r},\uparrow}-n_{\mathbf{r},\downarrow} \right)^2 \rangle,
\end{equation}
to probe the magnetic response.

Besides the spin-spin correlation functions, finite-temperature thermodynamic quantities offer complementary insights into magnetic correlation effects. Among them, the specific heat is particularly informative, since magnetic fluctuations are often manifested as characteristic peaks in its temperature dependence\cite{Paiva2001,Paiva2005,Medeiros-Silva2023}. To explore the thermodynamics in the kagome lattice, we compute the specific heat defined by
\begin{equation}
C(T) = \frac{1}{N} \frac{d\langle \mathcal{H} \rangle}{dT}.
\end{equation}
In the QMC simulations, the specific heat is obtained from the temperature derivative of the energy. The simulations provide energy data $E_n$ obtained at a set of inverse temperatures $\beta_n$, corresponding to discrete temperatures $T_n = 1/\beta_n$ $(n = 1,\ldots,N_T)$, where $N_T$ denotes the total number of simulated temperature points. Since numerical differentiation of discrete QMC data can lead to large fluctuations in $C(T)$, we perform an exponential fit to the energy data and then take the temperature derivative of the fitted function,
\begin{equation}
E_{\mathrm{fit}}(T) = a_0 + \sum_{l=1}^{M} a_l \exp(-\beta l \Delta),
\end{equation}
where the parameters $\Delta$ and $a_l$ are optimized by minimizing the cost function
\begin{equation}
\chi^{2} = \frac{1}{N_T} \sum_{n=1}^{N_T}
\frac{\left[ E_{\mathrm{fit}}(T_{n}) - E_{n} \right]^{2}}{(\delta E_{n})^{2}} .
\end{equation}
A cutoff of $M=6$ is employed. 
Further details of the fitting procedure and a representative example are provided in Appendix \ref{appendixA}.

\section{Results and discussion}\label{secresults}
\begin{figure}[tbp]
\centering
\includegraphics[width=\columnwidth]{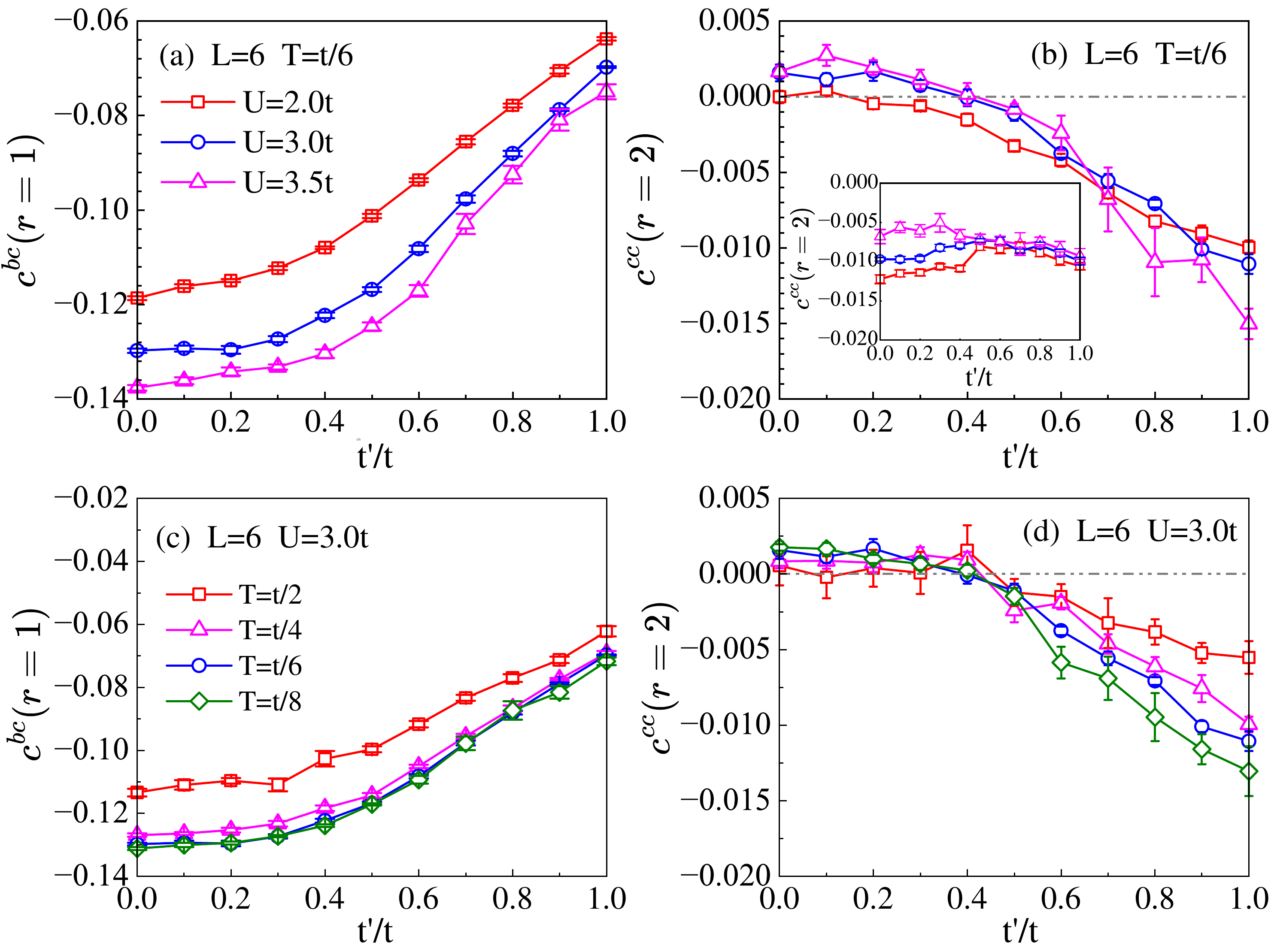}
\caption{(a) The spin-spin correlations $c^{bc}(r = 1)$ between the nearest-neighbor $b$ and $c$ sites and (b) the correlations $c^{cc}(r = 2)$ between the next-nearest-neighbor $c$ sites as a function of $t'/t$ for different $U$ at $L=6$ and $T=t/6$. 
In panel (b), the data correspond to the next-nearest-neighbor displacement $\mathbf{r}=(2,0)$; the inset shows correlations along $\mathbf{r}=(1,\sqrt{3})$.
(c) and (d) show $c^{bc}(r = 1)$ and $c^{cc}(r = 2)$ along $\mathbf{r}=(2,0)$ at different temperatures with $L=6$ and $U=3.0t$.
Error bars are not shown when they are smaller than the data points.}
\label{Fig2}
\end{figure}

To investigate the magnetic properties of the system shown in Fig.~\ref{Fig1}, we calculate the spin-spin correlations $c^{bc}(r = 1)$ between the nearest-neighbor $b$ and $c$ sites as a function of $t'/t$ for different interaction strengths $U$ at $L=6$ and $T=t/6$, as illustrated in Fig.~\ref{Fig2}(a). 
For all examined values of $U$, the nearest-neighbor correlations $c^{bc}(r=1)$ remain negative, reflecting that short-range antiferromagnetic correlations persist. 
The magnitude of the correlations $c^{bc}(r = 1)$ decreases as $t'/t$ increases, indicating that the increasing $t'/t$ suppresses the short-range antiferromagnetic correlations.
At fixed $t'/t$, the correlations $c^{bc}(r = 1)$ are enhanced with increasing $U$, indicating that stronger interaction $U$ promotes short-range antiferromagnetic correlations throughout the entire range of $t'/t$ considered. 
These observations may suggest a competition between electron correlations and frustration, where the frustration appears to have a stronger influence.

For longer-range part, the spin-spin correlations $c^{cc}(r = 2)$ between the next-nearest-neighbor $c$ sites exhibit a different trend.
There exists two inequivalent next-nearest-neighbor displacements $\mathbf{r}=(2,0)$ and $\mathbf{r}=(1,\sqrt{3})$ in the system. 
The correlations along $\mathbf{r}=(1,\sqrt{3})$, shown in the inset of Fig.~\ref{Fig2}(b), exhibit a weak antiferromagnetic signal which is nearly insensitive to $t'/t$. This behaviour may help explain the subtle low-temperature shoulder of the specific heat observed in some studies of the kagome lattice within certain parameter regimes \cite{Medeiros-Silva2023}, which may arise from the intrinsically weak antiferromagnetic correlations present in the system. 
However, the correlations along $\mathbf{r}=(2,0)$, shown in the main panel of Fig.~\ref{Fig2}(b), are more sensitive to $t'/t$.
The correlations undergo a sign change with increasing $t'/t$ and the corresponding crossover points shift to larger values as the interaction strength $U$ is enhanced. 
We also examine the temperature evolution of the spin correlations in Fig.~\ref{Fig2}(c)-(d). We find that the nearest-neighbor antiferromagnetic correlations build up at temperatures $T\ge t/3$. In comparison, the next-nearest-neighbor correlations along $\mathbf{r}=(2,0)$ emerge and develop at much lower temperatures, within the temperature range where the low-temperature specific-heat peak develops.
At $U = 3.0t$, the sign change of the next-nearest-neighbor correlations occurs around $t'/t \approx 0.3 \text{--} 0.4$.
As we will discuss later, the specific heat exhibits a low-temperature peak within this regime indicating a possible correlation between magnetic and thermodynamic responses. Further analyses will be provided in the discussion of the specific heat.

We have also examined the momentum-resolved spin structure factors $S(q)$. It is found that the maximum of $S(q)$ remains located at the $K$ point for all $t'/t$, corresponding to short-range $120^{\circ}$ coplanar antiferromagnetic correlations. Within the momentum-space resolution accessible in the present work, we do not find clear evidence for a significant change in the dominant magnetic configuration (see Appendix~\ref{appendixB} for details).

\begin{figure}[tbp]
\centering
\includegraphics[width=\columnwidth]{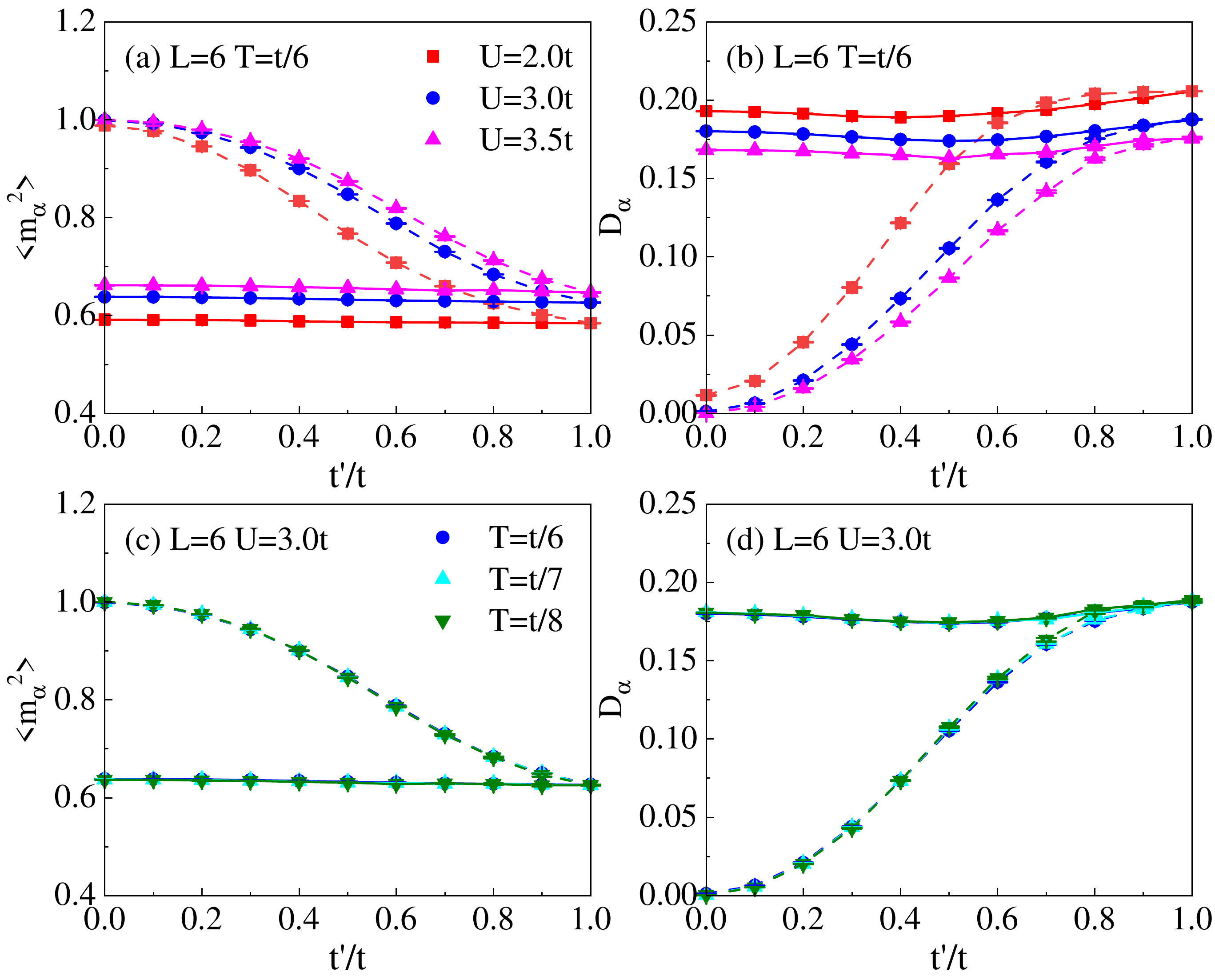}
\caption{
(a) Site-resolved local moment $\langle m_a^2 \rangle$ (solid lines) and $\langle m_d^2 \rangle$ (dashed lines) and (b) site-resolved double occupancy $D_a$ (solid lines) and $D_d$ (dashed lines) as a function of $t'/t$ for different $U$ at $L=6$ and $T=t/6$. (c) Site-resolved local moment and (d) double occupancy versus $t'/t$ for different temperatures $T$ at $L=6$ and $U=3.0t$. Error bars are not shown when they are smaller than the symbol size.
}
\label{Fig3}
\end{figure}

Figure \ref{Fig3}(a) shows the site-resolved local moment $\langle m_a^2 \rangle$ (for the equivalent $a$, $b$, and $c$ sites) and $\langle m_d^2 \rangle$ versus $t'/t$ for different $U$ at $L=6$ and $T=t/6$. 
The local moment is a measure of the degree of electron itinerancy. 
We find that $\langle m_d^2 \rangle$ decreases with increasing $t'/t$, indicating enhanced electron itinerancy. 
Meanwhile, $\langle m_a^2 \rangle$ shows little dependence on $t'/t$, suggesting that the charge fluctuations on the kagome sublattice ($a$, $b$, and $c$ sites) are weak at $T=t/6$. In this regime, the coupling between the kagome sublattice and d sites appears to be mainly mediated by spin correlations. 
The enhancement of the local moment with increasing $U$ is expected, since stronger on-site repulsion favors electron localization and hence local moment formation.
Figure \ref{Fig3}(b) shows the corresponding site-resolved double occupancy. We find $D_a$ is nearly independent of $t'/t$ whereas $D_d$ increases with increasing $t'/t$. This trend is consistent with the local moment results, since the local moment is related to the double occupancy through $\langle m^2 \rangle = \langle \left( n_{\mathbf{r},\uparrow}-n_{\mathbf{r},\downarrow} \right)^2 \rangle = \langle n_{\mathbf{r}}\rangle-2\langle n_{\mathbf{r},\uparrow}n_{\mathbf{r},\downarrow}\rangle=\langle n_{\mathbf{r}}\rangle-2\langle D\rangle$.

We further examine the temperature dependence of $\langle m_a^2 \rangle$ and $\langle m_d^2 \rangle$ in the low-temperature regime. As shown in Fig.~\ref{Fig3}(c), both $\langle m_a^2 \rangle$ and $\langle m_d^2 \rangle$ exhibit only a weak temperature dependence for $T=t/6$, $t/7$ and $t/8$, indicating that the local moment on all sites are already well developed and close to saturation at low temperatures. This suggests that the local moment formation mainly occurs at higher temperatures and contributes to the high-temperature specific heat peak, consistent with previous work \cite{Paiva2001}.
Correspondingly, both $D_a$ and $D_d$ show only a weak temperature dependence shown in Fig.~\ref{Fig3}(d).

\begin{figure}[tbp]
\centering
\includegraphics[width=\columnwidth]{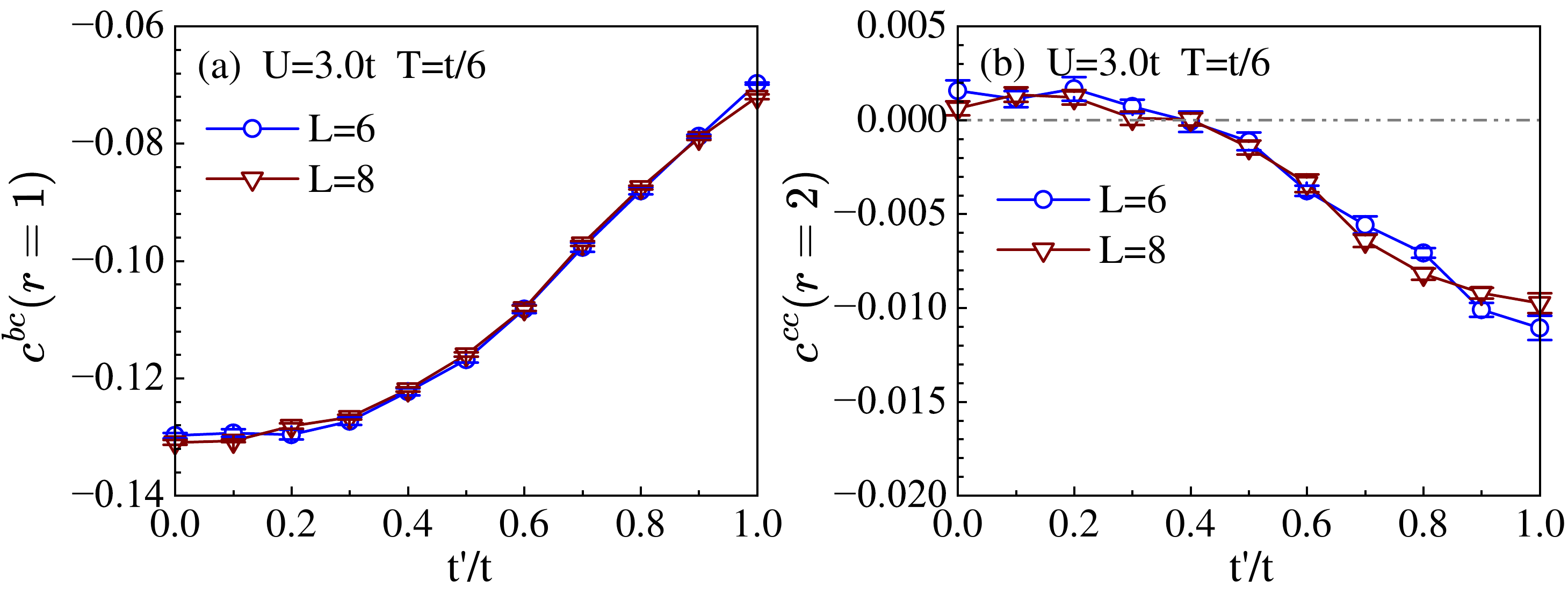}
\caption{(a) The spin-spin correlations $c^{bc}(r = 1)$ and (b) the correlations $c^{cc}(r = 2)$ along $\boldsymbol{r}=(2,0)$ for lattice sizes $L=6$ and $L=8$ at fixed $U=3.0t$ and $T=t/6$. Error bars are not shown when they are smaller than the data points.}
\label{Fig4}
\end{figure}

To assess possible finite-size effects, we have examined selected parameters and observables on different lattice sizes.
Figure \ref{Fig4} displays the correlations $c^{bc}(r = 1)$ and $c^{cc}(r = 2)$ as a function of $t'/t$ for $L = 6$ and $L = 8$, with fixed $U = 3.0t$ and $T=t/6$.
It is evident that these correlations show no significant dependence on the system size. In particular, the value of $t'/t$ at which the sign change occurs remains nearly unchanged as $L$ increases.
These results confirm that our calculations based on $L = 6$ are well-converged and reliable.
In strongly geometrically frustrated systems, it is argued that long-range magnetic order at zero or very low temperatures may exhibit pronounced sensitivity to system size and boundary conditions \cite{Cookmeyer2021}. At the finite temperature considered here, we do not observe significant magnetic correlations that are sensitive to system size, which may be attributed to the limited temperature range accessible in our simulations.

\begin{figure}[tbp]
\centering
\includegraphics[width=\columnwidth]{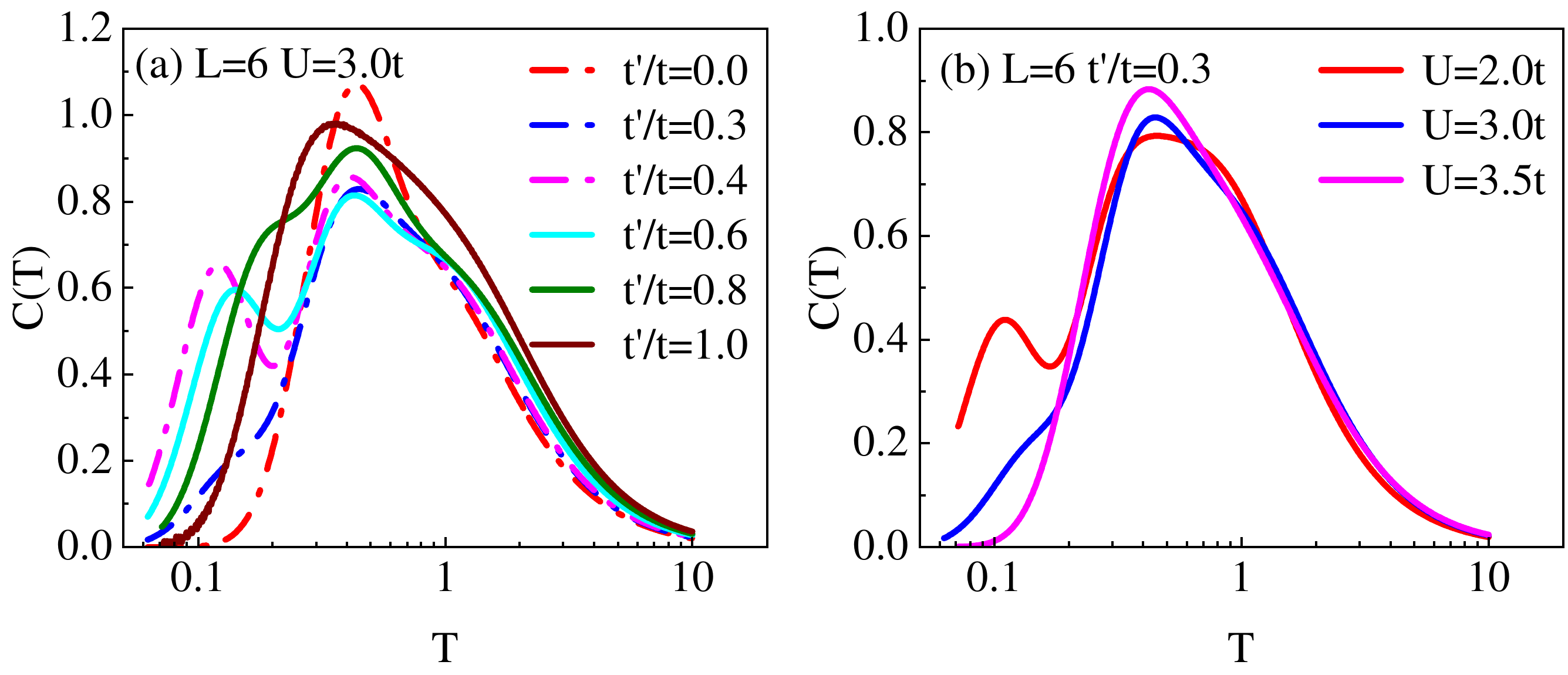}
\caption{The specific heat $C(T)$ as a function of temperature $T$ (a) for different values of $t'/t$  at $L=6$ and $U=3.0t$ and (b) for different values of $U$ at fixed $t'/t=0.3$. The specific heat is obtained from the exponential fit and is shown in a log-linear scale.}
\label{Fig5}
\end{figure}

We now turn to the thermodynamic properties of the system. 
Figure \ref{Fig5}(a) presents the specific heat $C(T)$ as a function of temperature $T$ for different values of $t'/t$ at $U=3.0t$.
In the limit of $t'/t = 0$, the system consists of a kagome lattice with a set of decoupled $d$ sites.
In this case, the specific heat is primarily determined by the kagome lattice.
Our result at $t'/t=0$ is qualitatively consistent with previous QMC studies \cite{Medeiros-Silva2023} that report a broad high-temperature peak in the specific heat of the kagome lattice. 
This high-temperature feature is related to the formation of local moment.
Notable changes occur at low temperatures as $t'/t$ increases. 
At $t'/t = 0.3$ a soft shoulder emerges and a low-temperature peak becomes clearly visible at $t'/t = 0.4$. 
The appearance of this low-temperature peak often indicates low-lying collective spin-wave excitations, and the double-peak structure suggests strong spin-spin correlations \cite{Medeiros-Silva2023}. 
We associate this feature with the enhancement of the next-nearest-neighbor antiferromagnetic correlations $c^{cc}(r = 2)\le0$ with increasing $t'/t$. 
We note that the kagome lattice hosts only weak next-nearest-neighbor antiferromagnetic correlations, which are insufficient to produce a pronounced low-temperature structure in the specific heat and may give rise only to a subtle shoulder feature. 
As $t'/t$ increases, the correlations $c^{cc}(r = 2)$ are significantly enhanced. 
In particular, for $t'/t \approx 0.3 \text{--} 0.4$, the antiferromagnetic correlations $c^{cc}(r = 2)$ along $\boldsymbol{r}=(2,0)$ emerge and are enhanced significantly, directly contributing to the emergence of a pronounced low-temperature peak in the specific heat.
As $t'/t$ increases further, Figure \ref{Fig5}(a) shows that the low-temperature peak persists and gradually shifts toward higher temperatures. In the limit of $t'/t = 1$, it eventually merges with the high-temperature peak, forming a broad single maximum. 
It has been suggested that the low-temperature and high-temperature peaks of the specific heat correspond to spin and charge energy scales respectively.  
Accordingly, the emergence of the low-temperature peak signals the development of a spin-related energy scale, consistent with the enhancement of next-nearest-neighbor antiferromagnetic correlations discussed above.
These features provide insight into how the spin and charge energy scales affect the thermodynamic behavior of the kagome lattice and related frustrated systems.

We further examine the influence of the interaction strength $U$ on the specific heat.
Fig.~\ref{Fig5}(b) displays the low-temperature features of the specific heat at a fixed $t'/t = 0.3$ for different values of $U$. 
A discernible low-temperature feature is already present at $U = 2.0t$, but becomes progressively suppressed with increasing $U$ and disappears at $U = 3.5t$.
This trend is consistent with the evolution of the next-nearest-neighbor correlations shown in Fig.~\ref{Fig2}(b). At fixed $t'/t=0.3$, the antiferromagnetic correlations $c^{cc}(r = 2)$ are developed at $U = 2.0t$ but remain weak at $U = 3.5t$.
The correlated evolution of the specific heat and spin correlations supports a close connection between magnetic correlations and the low-temperature thermodynamic response.

\begin{figure}[t]
\centering
\includegraphics[width=\columnwidth]{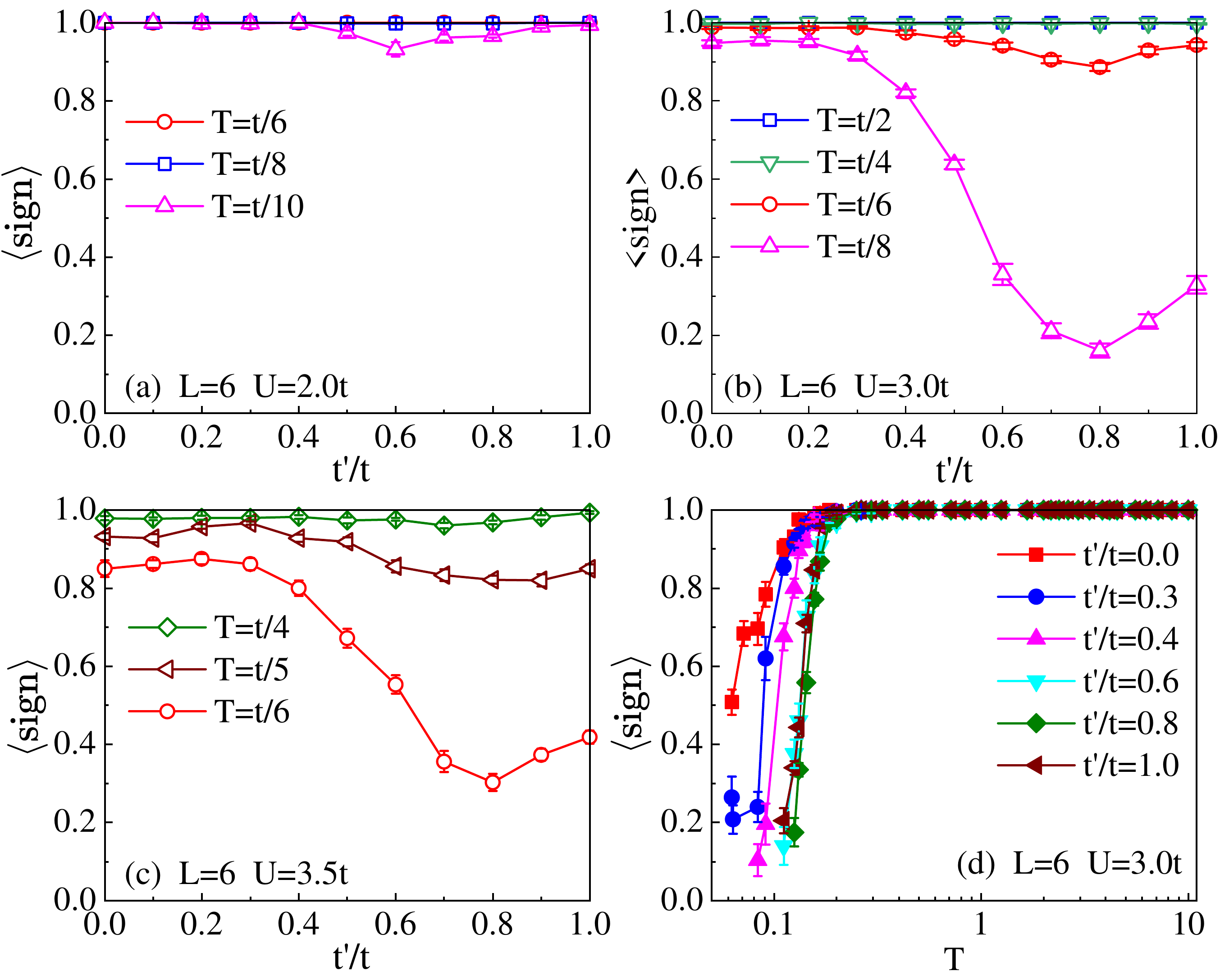}
\caption{The average sign $\langle \text{sign} \rangle$ of the fermionic determinant as a function of $t'/t$ is shown. Each of the panels corresponds to a fixed value of $U$, and every curve within a panel represents at different temperatures $T$, as indicated in (a)-(c). Panel (d) shows $\langle \text{sign} \rangle$ as a function of $T$ for different $t'/t$ at $L=6$ and $U=3.0t$.  Error bars are not shown when they are smaller than the data points.}
\label{Fig6}
\end{figure}

We proceed to examine the infamous sign problem, which is a significant challenge in the kagome lattice due to its strong frustration.
Fig.~\ref{Fig6}(a)-(c) shows the behavior of $\langle \text{sign} \rangle$ as a function of $t'/t$ for different values of $U$ and temperature $T$. 
It is found that larger $U$ and lower $T$ tend to intensify the sign problem, which is in agreement with previous calculations in kagome systems.
For $U=2.0t$ we are able to reach low temperatures in the simulation, as the sign problem remains manageable when temperature is lowered to $T=t/10$. 
However, the sign problem intensifies rapidly as $U$ increases. 
Since the physical properties of interest generally require $T\leq t/5$, we limit our study to $U \leq 3.5t$.
We also attempt simulations at $U=4.0t$, but find that $\langle \text{sign} \rangle$ drops to nearly zero for $t'/t \geq 0.6$, indicating severe sign problems. 
We hope that future work may enable systematic studies at stronger interactions.
In Fig.~\ref{Fig6}(d), we show the temperature dependence of $\langle \text{sign} \rangle$ for different $t'/t$ at $L=6$ and $U=3.0t$. 
As the temperature decreases, the sign problem becomes more severe, with $\langle \text{sign} \rangle$ dropping rapidly.
The dips in $\langle \text{sign} \rangle$ become deeper with increasing $t'/t$, indicating the complex interplay between frustration and electron correlations.
Given that the bandwidth of the system is $W=6t$, an interaction strength of $U = 3.0t$ already falls within the intermediate coupling regime, where the sign problem is expected to be significant. In this region, we focus on data where $\langle \text{sign} \rangle \gtrsim 0.1$.
Furthermore, previous studies have suggested that the average sign may serve as an additional observable to locate quantum critical points \cite{Mondaini2022}.

\section{Conclusions}\label{secconclusions}
In this work, we investigate the Hubbard model on the kagome lattice with an additional hopping $t'/t$ that continuously interpolates between the kagome and triangular lattices. 
Using determinant quantum Monte Carlo simulations, we explore the evolution of magnetic and thermodynamic responses. 
Our results show that increasing $t'/t$ suppresses the nearest-neighbor antiferromagnetic correlations.
Meanwhile, the next-nearest-neighbor antiferromagnetic correlations are enhanced and closely associated with the emergence of a pronounced low-temperature peak in the specific heat.
Finite-size analysis further confirms the robustness of these conclusions.
These findings demonstrate that the additional hopping $t'/t$ effectively tunes the competition between geometric frustration and electronic correlations in the kagome lattice, giving rise to emergent low-energy spin excitations.
Moreover, the analysis of the fermionic sign problem further clarifies its parameter dependence and provides guidance for simulations in stronger-coupling regimes.

Our results may offer a perspective for interpreting the anomalous low-temperature specific heat features reported in experiments. 
In \ce{Dy3Sb3Zn2O14} with significant Dy/Zn site mixing, a pronounced anomaly in the specific heat at $T \approx 0.35 \mathrm{K}$ has been observed and is generally attributed to long-range interactions such as dipolar coupling or next-nearest-neighbor exchange. 
By introducing $t'/t$, our model continuously interpolates between the kagome and triangular lattices, analogously mimicking the impact of site mixing in \ce{Dy3Sb3Zn2O14}, which perturbs the ideal kagome lattice and introduces a geometry resembling a diluted two-dimensional triangular lattice or an intermediate kagome-pyrochlore structure. 
We propose that finite $t'/t$ generates spin correlations extending beyond the nearest-neighbor sites, leading to an unusual low-temperature feature in the specific heat.
Although the simulations have not yet reached the extremely low temperatures accessed experimentally, the results remain relevant because they capture the emergence of longer-range spin-spin correlations and energy scales that are qualitatively consistent with the low energy features in experiments. This behavior indicates the presence of an additional spin-related energy scale, offering insight into the multiple energy scales observed in kagome systems and other frustrated correlated materials.

\appendix
\section{Specific heat calculation and fitting procedure}\label{appendixA}

In our DQMC simulations, the specific heat is obtained from the temperature derivative of the energy,
\begin{equation}
C(T) = \frac{1}{N}\frac{\mathrm{d}\langle H \rangle}{\mathrm{d}T}.
\end{equation}
Since the energy is obtained at discrete temperature points with statistical errors, we extract the specific heat by fitting the energy data to an exponential function and taking the temperature derivative of the fitted energy. 
Similar fitting procedures for extracting the specific heat from QMC energy data have been employed in previous studies~\cite{Paiva2001,Medeiros-Silva2023}.

\begin{figure}[tbp]
\centering
\includegraphics[width=\columnwidth]{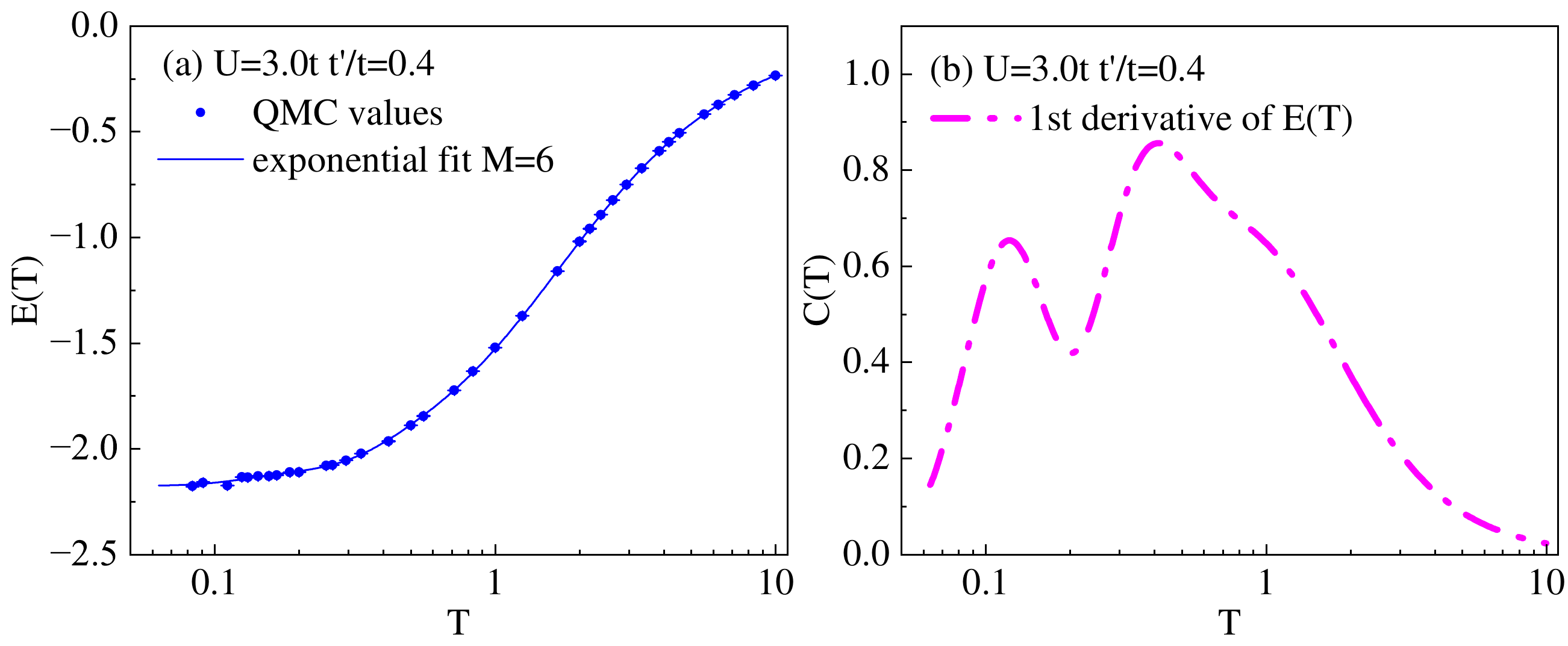}
\caption{Representative example of the exponential fitting procedure applied to the QMC energy data $E_n(T)$ for $U=3.0t$ and $t'/t=0.4$, together with the resulting specific heat obtained from the temperature derivative of the fitted energy.}
\label{Fig7}
\end{figure}

Figure~\ref{Fig7} shows a representative example of the fitting procedure for $U = 3.0t$ and $t'/t = 0.4$. The fitted energy closely follows the QMC data over the temperature range considered, and the specific heat is obtained from the temperature derivative of the fitted energy.

\section{Momentum-resolved spin structure factor}
\label{appendixB}
In addition to the real-space spin correlations discussed in the main text, we compute the momentum-resolved spin structure factor to further characterize the magnetic correlations by identifying the dominant fluctuation wave vectors. The spin structure factor is defined as
\begin{equation}
S(\mathbf{q}) = \frac{1}{N} \sum_{i,j} e^{i \mathbf{q} \cdot (\mathbf{r}_i - \mathbf{r}_j)} \langle \mathbf{S}_i \cdot \mathbf{S}_j \rangle,
\end{equation}
where $N$ is the total number of lattice sites and $\mathbf{r}_i$ denotes the position of site $i$.

\begin{figure}[h]
\centering
\includegraphics[width=0.6\columnwidth]{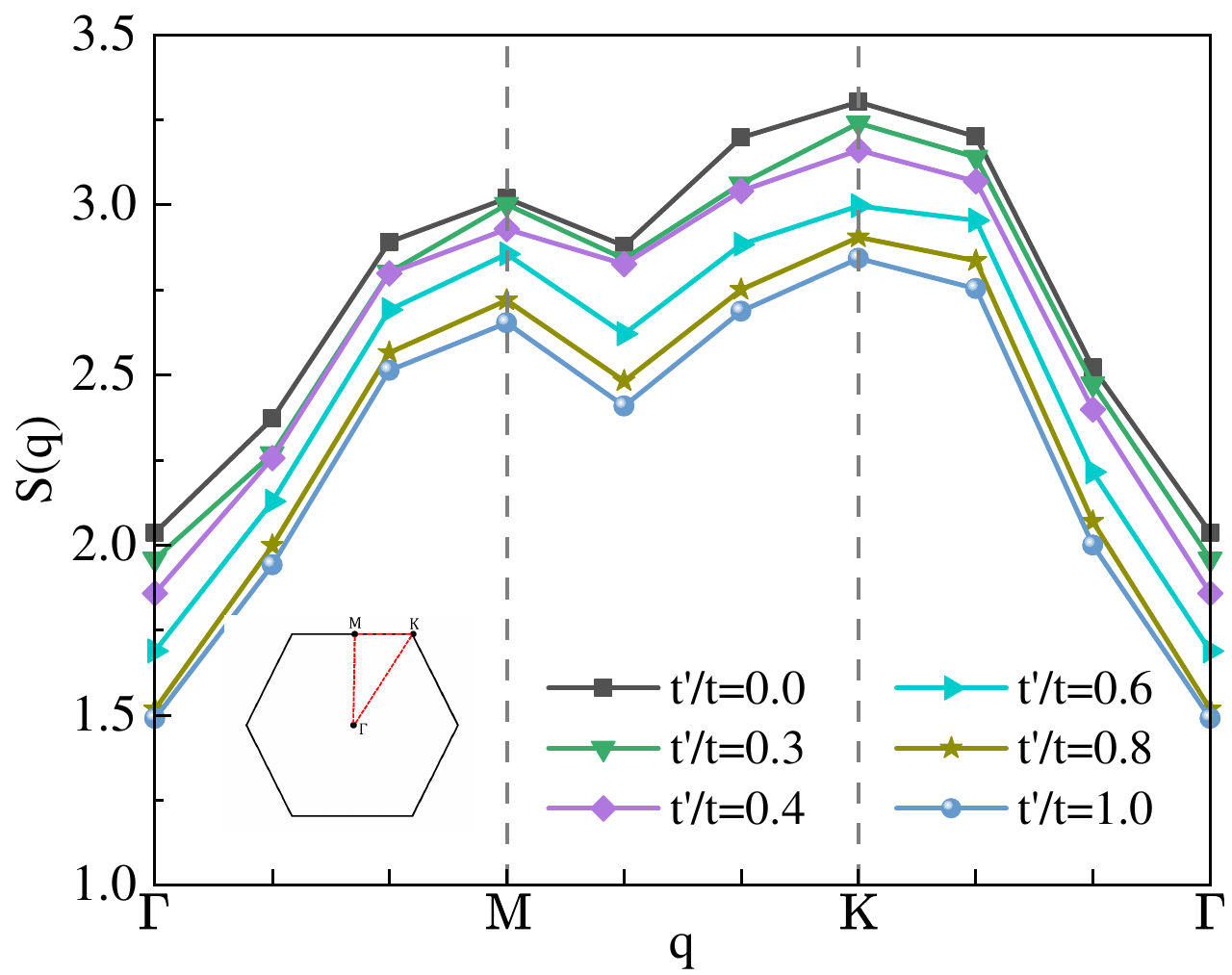}
\caption{Momentum-resolved spin structure factor $S(\mathbf{q})$ along a high-symmetry path in the Brillouin zone for representative values of $t'/t$.}
\label{Fig8}
\end{figure}

Figure~\ref{Fig8} shows $S(\mathbf{q})$ along a representative high-symmetry path in the Brillouin zone for different values of $t'/t$. We find that the maximum of $S(\mathbf{q})$ remains located at the $K$ point over the entire range of $t'/t$ considered. This indicates that the dominant magnetic correlations retain the $K$-type character upon tuning $t'/t$, corresponding to short-range $120^{\circ}$ coplanar antiferromagnetic correlations.

At the same time, the overall magnitude of $S(\mathbf{q})$ decreases gradually as $t'/t$ increases. This trend is consistent with the reduction of the local moment discussed in the main text and reflects the enhancement of electron itinerancy induced by the additional hopping $t'$. 
Within the momentum-space resolution accessible in the present simulations, we do not find clear evidence for a qualitative change in the dominant magnetic configuration. 

\section*{Acknowledgments}
This work was supported by National Natural Science Foundation of China (Grants No. 12474218) and Beijing Natural Science Foundation (Grants No. 1242022 and No. 1252022). The numerical simulations in this work were performed at the HSCC of Beijing Normal University.

\section*{DATA AVAILABILITY}
The data that support the findings of this article are openly available~\cite{jia2026data}.

\bibliography{reference}

\end{document}